\documentclass{article}
\usepackage{fullpage}
\usepackage{amsfonts,amssymb}
\usepackage{amsthm}
\usepackage{tikz}
\usetikzlibrary{calc}
\usetikzlibrary{arrows}
\usepackage{tikz-3dplot}
\usepackage{color} 
\usepackage[fleqn]{amsmath}
\usepackage{amsthm}
\usepackage{amssymb}
\usepackage{amsfonts}
\usepackage{bbm}
\usepackage{epsfig}
\usepackage{times}
\usepackage{hyperref}
\usepackage{bm}
\usepackage{float} 
\usepackage{appendix} 
\usepackage{lscape}
\usepackage{cite}
\usepackage{mathrsfs}
\usepackage{appendix}
\usepackage{setspace}
\usepackage{mathtools}

\theoremstyle{definition}
\newtheorem{definition}{Definition}[section]
\newtheorem{lemma}[definition]{Lemma}
\newtheorem{corollary}[definition]{corollary}

\newtheorem{theorem}{theorem}

\begin{document}
 
\title{The Operational Choi-Jamio\l{}kowski  Isomorphism}

\author{Emily Adlam \footnote{ 
		Basic Research Community for Physics, Leipzig, Germany; eadlam90@gmail.com }}  

\maketitle
  
In this article, I use an operational formulation of the Choi-Jamio\l{}kowski  isomorphism to explore an approach to quantum mechanics in which the state is not the fundamental object. 
	I first situate this project in the context of generalized probabilistic theories and argue that this framework may be understood as a means of drawing conclusions about the intratheoretic causal structure of quantum mechanics which are independent of any specific ontological picture.  I then give an operational formulation of the Choi-Jamio\l{}kowski  isomorphism and show that in an operational theory which exhibits this isomorphism, several features of the theory which are usually regarded as properties of the quantum state can be derived from constraints on non-local correlations.  This demonstrates that there is no need to postulate states to be the bearers of these  properties, since they can be understood as consequences of a fundamental equivalence between multipartite and temporal correlations.

 \section{Introduction}
 
 In the standard mathematical formulation of quantum mechanics, the quantum state is the most fundamental object. However, there exists a longstanding debate about whether we should regard the quantum state as an element of reality\cite{Hardy2013, Quanta22}. A recent result known as the PBR theorem purported to settle this debate once and for all by showing that no interpretation of quantum mechanics where the quantum state is not an element of reality can reproduce all of the theory's empirical predictions\cite{PBR}. But in order to prove the PBR theorem it is assumed that information about the past must be carried into the future by a mediating state,  so in fact the setup for the  theorem takes it for granted that the fundamental object in quantum mechanics is some sort of temporally evolving state, whether or not that state turns out to be the \emph{quantum} state. 
 
 The assumption that temporal correlations must be mediated by states is seldom seriously questioned, but it is in fact in tension with certain elements of modern physics\cite{Adlamspooky}, and this suggests that we ought to take seriously the possibility of re-imagining quantum mechanics with an ontology that does not employ states at all. Quantum states have certain fundamental temporal properties (for example, they can't be broadcasted, they are contextual,  they exhibit interference effects, and they are subject to uncertainty relations) and a literal reading of the formalism of quantum mechanics would suggest that the characteristic features of the theory are derived from these properties of the quantum state. Thus as a first step towards a stateless ontology, we should determine whether it is necessary to postulate some sort of state to be the bearer of these properties, or whether they can be accounted for in another way. 
 
 In recent years there have been moves within the quantum foundations community toward a type of research where quantum mechanics is placed in the context of a space of generalized probabilistic theories (GPTs), with the aim of showing that certain features of quantum mechanics can be derived from some chosen axioms  \cite{ Masanes_2011, CDK, Hardyreasonable, grinbaum2003element, fuchs2002quantum, Rohrlich, Pawlowski, Toner,  Bub}. However, most of these projects use axioms which employ a concept of `state,' and hence they don't offer an obvious route to a stateless ontology. In this paper, I employ the methodology of the generalized probabilistic theories framework, but I develop an alternative approach which is explicitly geared towards eliminating the concept of state.

 I begin in section \ref{IC} by elaborating on the rationale for studying generalized probabilistic theories.  Then in section \ref{CJ} I set out an operational form of the  Choi-Jamio\l{}kowski isomorphism, and demonstrate that this operational isomorphism can be used to derive  temporal properties of states from constraints on the types of non-local correlations that are permitted within the theory. This  demonstrates that in fact we don't require a state to be the bearer of the aforementioned temporal properties, and thus paves the way for a new way of thinking about the intratheoretic causal structure of quantum mechanics where states are not fundamental objects.

 \section{Intratheoretic Causal Structure \label{IC}}
 
 \subsection{Ontology}
 
 The issue of ontology is a vexed one in physics. It is a well-established fact that ontology is always underdetermined by the empirical results, and therefore no amount of experimentation can give us certainty about the nature of the unobservable entities associated with our theory\cite{sep-scientific-underdetermination}. Furthermore, when one theory is replaced by another, it is often the case that the theoretical objects postulated by the original theory do not feature in the new theory, and thus unless we believe our current theory is the correct final theory of reality we should probably assume that the theoretical objects posited by our current theory will not feature in future theories\cite{sep-structural-realism}. Considerations such as these have led many physicists to adopt variants of anti-realism, positivism or empiricism, holding that physics should be concerned only with empirical predictions and should be silent on questions of ontology\cite{sep-scientific-realism,van1980scientific}. 
 
 However, 
 it is important to recognise that the ontology we associated with our theories affects the way in which science progresses. For example, recall that classical physics can be written either in the standard Newtonian form, where we start with initial positions and calculate trajectories from those positions, or in a Lagrangian form where we define an 'action' over an entire history and find the trajectory which optimizes the action \cite{Wharton, brizard2008introduction}. Historically most physicists have considered the Newtonian formulation of classical physics to be more fundamental than the Lagrangian one, and in this context it was natural for the founders of quantum mechanics to formulate the new theory in terms of a state evolving forwards in time, even though  we now know that an alternative Lagrangian formulation in terms of path integrals exists\cite{Wharton, feynman2010quantum}. Quantum mechanics might look very different (and indeed might well be less difficult to 'interpret') had we taken the Lagrangian formulation seriously from the start.
 
 Moreover, simply becoming an operationalist and refusing to talk about ontology doesn't confer immunity to these sorts of biases, because ontological assumptions  are baked into the way in which we think about and do physics\cite{Wharton, spekkens2012paradigm, Adlamspooky}, and so those who decline to think seriously about ontology are effectively choosing to uphold the status quo ontological assumptions - which, in the case of quantum mechanics, entails upholding the assumption of an ontology based on states.

 Thus reconsidering the ontology associated with our theories can be a useful way of flushing out unjustified assumptions and ultimately coming up with new physics.  
 However, in adopting this strategy we risk getting caught up in the details of some specific ontology, which due to the problem of underdetermination will inevitably not be correct in every particular. Ideally, what we would like to do is extract the significant \emph{structural} content of various different ontologies and discard the specifics, which don't really contribute to the project of coming up with new physics and are essentially just window-dressing. This is exactly what the framework of generalized probabilistic theories allows us to do\cite{ Masanes_2011, CDK, Hardyreasonable, grinbaum2003element, fuchs2002quantum, Rohrlich, Pawlowski, Toner,  Bub}: it offers a middle way between realism and empiricism, where we impose hierarchies within our theories by regarding certain features of the theory as consequences of other features, but do not actually specify the details of the ontology underlying the hierarchy. 
 
 \subsection{Intratheoretic Causation}

 Hierarchical structures of this kind occur in many places throughout the sciences.\footnote{I observe in passing that some philosophers have questioned the coherence of imposing a hierarchy of fundamentality on nature. I will not engage with this debate here, but will instead observe that whatever its metaphysical status, the `hierarchy of fundamentality' picture is instrumentally useful in the actual practice of science as a means of rethinking the structural assumptions that go into our theories.}. They are most straightforward in theories which come with an ontology 
 attached, because in such cases a hierarchy of fundamentality can be read directly off the ontology. For example, the ontology of classical physics postulates objects which move in inertial reference frames and are subject to certain mechanical forces, but centrifugal force is not one of those mechanical forces - rather it is is a `fictitious' force which arises when we try to describe an object in a reference frame which is accelerating relative to an inertial reference frame\cite{K_chemann_2020}. This ontology encodes a hierarchy where the centrifugal forces are to be understood as consequences of the more fundamental laws governing motion relative to inertial frames. Indeed, the hierarchy licences counterfactual assertions about the relationship between these features: if the laws governing motion relative to inertial frames had been relevantly different, the centrifugal forces would also have been different. Thus we observe that these hierarchies of fundamentality are roughly analogous to causal structures, with `causation' to be understood in the counterfactual sense: in a given theory, feature A causes feature B if feature B would not have obtained if A had not obtained.\footnote{There are many subtleties that have to be dealt with in any counterfactual account of causation; I will not deal with any of them here, but refer to ref \cite{sep-causation-counterfactual} for a general discussion of counterfactual approaches to causation. It is likely that further subtleties would arise in the application of this approach to intratheoretic causation; I will return to these in forthcoming work.} Therefore I will henceforth refer to these hierarchies as `intratheoretic causal structures' though I emphasize that the word `causal' is here being used in a specialised sense, since the relata of the putative causal relation are not immanent. 
 
 We are already quite familiar with the  case where the intratheoretic causal structure can be read off the ontology. But what if there is no general consensus regarding the ontology of a theory? Quantum mechanics is one such case – to this day it is not even clear whether quantum mechanics requires the existence of the wavefunction\cite{Leiferreview}, let alone other proposed features such as spontaneous collapses\cite{Ghirardi1995}, de Broglie-Bohm particles\cite{holland1995quantum} or `many-worlds.'\cite{Wallace} And as a result the intratheoretic causal structure of quantum mechanics is much less transparent - witness the continuing controversy over the reality of the quantum state, which can be understood as a discussion about whether we should adopt an intratheoretic causal structure in which the characteristic features of quantum mechanics are derived from the properties of the quantum state, or whether we should take it that the quantum state itself is derived from some deeper properties. 
 
 Obviously it is open to us to resolve these questions by adopting an interpretation of quantum mechanics and drawing the conclusions implied by its ontology. For example, the Everett interpretation tells us that the quantum state is indeed the fundamental object of quantum mechanics\cite{Wallace}, whereas Quantum Bayesianism suggests that the quantum state is simply a description of an agent's degrees of belief about the outcomes of measurements\cite{FuchsMermin}. But the study of generalized probabilistic theories offers an alternative, allowing us to  draw conclusions about intratheoretic causal structure which are independent of any specific ontological assumptions. Research of this kind works by placing quantum mechanics in the context of a wider space of GPTs so that we can meaningfully consider counterfactual questions about how changing one aspect of a theory might lead to other changes\cite{ Masanes_2011, CDK, Hardyreasonable, grinbaum2003element, fuchs2002quantum, Rohrlich, Pawlowski, Toner,   Bub}. This often involves proving  statements of the form: `any generalized probabilistic theory which has feature $X$ must also have feature $Y$.' For example, ref \cite{Toner} proves that any GPT which obeys no-signalling must also obey a form of monogamy of correlations, thus demonstrating the existence of a (putatively causal) relationship between these features which holds regardless of what the underlying ontology might be. The framework of generalized probabilistic theories therefore provides exactly the formalism needed for us to study the counterfactual reasoning associated with intratheoretic causal structure in a way that does not depend on any particular ontology.

 Of course, showing that certain features of quantum mechanics can be derived from other features in this way doesn't actually \emph{prove} anything about the intratheoretic causal structure of quantum mechanics, because there exists no unique axiomatization of quantum mechanics and each different axiomatization will suggest different intratheoretic causal relations. But that is just to restate the old problem of underdetermination of theory by data, and a similar response can be made: the point is not necessarily to know for certain, but to understand the implications of different possible causal structures as a way of exploring alternative ways forward for physics. One need not believe that any particular causal structure is the correct one, or indeed that a unique correct causal structure even exists, to see the value of the exercise.

 \subsection{Structural Realism}
 
 I have suggested that one motivation for research in the space of generalized probabilistic theories is the idea that we don't have much hope of coming to know the true ontology of our theories, so instead of focusing on the details of specific ontologies we should instead think in terms of intratheoretic causal structures, which plausibly we could  get right even if the true ontology remains epistemically inaccessible. This suggestion clearly has common ground  with the philosophical view known as structural realism\cite{sep-structural-realism}, which proposes that since the specific ontologies of our theories will likely be discarded when we move to a new theory, we should not be epistemically committed to specific ontologies but instead commit to  the structural relations implied by the ontologies of the theories, since these relations are often maintained when we move to a new theory.   However, the relations to which structural realism refers are usually instantiated by specific objects in the world, whereas the intratheoretic relations that I discuss here hold between parts of a theory, so the domain of application of this form of structuralism is importantly different.

 \section{The Choi-Jamio\l{}kowski Isomorphism \label{CJ}}

 In the remainder of this article, I will elaborate on an intratheoretic causal structure for quantum mechanics where the quantum state is not the fundamental object. My approach is based on the Choi-Jamio\l{}kowski isomorphism, which describes the mathematical correspondence between quantum channels and entangled bipartite states \cite{Jiang}. The Choi-Jamio\l{}kowski isomorphism can be derived from the standard  mathematical formalism of quantum mechanics, and the fact that it is usually taught in this way encourages physicists to regard it as a consequence of the properties of the quantum state. In particular, it is often interpreted by appeal to an operational procedure known as  `noisy gate teleportation' in which agents $A$ and $B$ share entanglement and agent $A$ is in possession of a system in an unknown state $\rho$ and the aim is for $B$ to end up with a system in the state $T \rho$ where $T$ is some fixed transformation; this cannot usually be achieved with perfect certainty, but insofar as it is possible it works because of the Choi-Jamio\l{}kowski isomorphism\cite{LSpekkens, Gottesman}. This presentation implies that the isomorphism is just one of the many surprising operational consequences which follows from the properties of the quantum state, so it is of no special foundational interest.

 But consider: what the isomorphism actually tells us (roughly speaking), is that in quantum mechanics, the set of possible multipartite correlations exhibited by entangled states is equivalent to the set of possible temporal correlations exhibited by sequences of measurements on a single system over time. This equivalence has the air of a fact about ontology. Indeed, I would argue that and indeed, any choice for the ontology of quantum mechanics which failed to reflect such a striking equivalence would surely be suspect on the grounds of exhibiting `asymmetries which do not appear to be inherent in the phenomena.'\footnote{This phrasing is of course borrowed from ref \cite{Einstein1905}}. Thus in this paper I will suppose that the Choi-Jamio\l{}kowski isomorphism is indeed a deep fact about the underlying ontology of quantum mechanics, and therefore causally prior to most other features of the theory.

 In accordance with the discussion of section \ref{IC}, rather than proposing a specific ontology which has this feature, I will proceed directly to an investigation of the consequences of this conjecture for the intratheoretic causal relations in quantum mechanics. In section \ref{CJ} I give an operational formulation of the Choi-Jamio\l{}kowski isomorphism, which relates measurements on different parts of a multipartite system to sequences of measurements on the same system and vice versa. In sections \ref{no-broadcasting}, \ref{contextualitysec} and \ref{uncertainty} I use it to prove results of the form `Any generalized probabilistic theory which obeys the operational Choi-Jamio\l{}kowski isomorphism, and which has property A, must also have property B,' where A is a constraint on multipartite measurements and B is a constraint on sequences of measurements, as shown in the table below. That is, I show that several features of quantum mechanics which are typically interpreted as temporal properties of the quantum state could in fact be derived from the properties of some more general entity which is the substratum for both multipartite states and time evolution.

 \vspace{3mm}

 \hspace{-15mm}	\begin{tabular}{ l | l } 
 	\label{corr}
 	
 	\textbf{	Property} & \textbf{Implies} \\ \hline
 	
 	Strong monogamy of correlations and Bell nonlocality & No-broadcasting  and quantum interference
 	
 	\\ \hline
 	No-signalling and Bell nonlocality &    Preparation contextuality \\ \hline 
 	Information causality & Fine-grained uncertainty relations \\ \hline 
 	\vspace{2mm} 
 	
 \end{tabular}
 
 Each of the properties in this table will be given an operational definition in subsequent sections. Note that the concept of `operational state' will feature in the definition of some of the properties on the right-hand side of the table, since they have been chosen specifically as properties which are normally attached to states. However, crucially, the concept of state is not used in the operational definitions of the concepts on the left-hand side, or in the operational Choi-Jamio\l{}kowski isomorphism. Thus as promised, these results demonstrate features that appear to be properties of states can in fact be understood as consequences of features of the theory which do not depend in any way on a concept of state.

 \subsection{Related Work}
 
 Although the mainstream literature on quantum mechanics is still very dominated by state-based approaches, in recent years a variety of interesting work has been done on non-standard temporal pictures. For example, there has been growing interest in retrocausal approaches to quantum theory\cite{Goldstein_2003, PRICE_1994, Priceretro}, including a proof by Pusey and Leifer demonstrating that if quantum mechanics obeys a certain sort of time-symmetry then it must exhibit retrocausality\cite{PuseyLeifer} and a model due to Wharton which suggests a natural resolution to the quantum
 reality problem using the “all-at-once”-style analysis of action principles\cite{Whartoninformation}. Elsewhere,  Oreskhov and Cerf\cite{Oreshkov2} have set out the process matrix formalism, which allows us to generalize the framework of operational theories in a way that does not depend on a predefined time or causal structure, thus giving us the mathematical resources to deal with theories that might contain indefinite causal order, causal loops or other structures that don't fit into our familiar notions of time and causality. Similarly,  Shrapnel and Costa have used the process matrix formalism to set out a generalisation of the ontological models approach which does not assume that information must be carried through time by a mediating state, and have used this generalisation to demonstrate that even without mediating states quantum mechanics must still exhibit a generalized form of contextuality\cite{Shrapnel_2018}. Thus the ideas set out here add to a growing body of work on the ways in which quantum mechanics could be embedded into global and temporally non-local structures. 
 
It has been recognised that the Choi-Jamiolkowski isomorphism is likely to play an important role in such non-standard temporal pictures - in particular, the process matrix is defined using the Choi-Jamiolkowski representations of the relevant CP maps\cite{Oreshkov2}. The process matrix formalism therefore assigns to the isomorphism an implicit ontological significance, and so the results presented here can be understood as complementary to that line of research, since I have made the ontological significance of the isomorphism explicit and used an operational formulation to explore its consequences for intratheoretic causal structure. A variety of other authors have also suggested that the isomorphism should be understood in ontological terms, so for example it is noted in \cite{Verstraetechannels}  that the Choi-Jamio\l{}kowski representation for quantum operations \emph{`gives a nice way of unifying statics and dynamics in one framework: the future is entangled (or at least correlated) with the past,'} and likewise,  ref \cite{AharonovPopescuTollaksen} puts forward a theoretical model in which \emph{ `one particle at N times is ... equivalent to N particles at one time.' }

Finally, a different version of an `operational  Choi-Jamio\l{}kowski isomorphism' was put forward  in ref \cite{Chiribella}, where it is shown that in any operational theory which a) is causal, and b) has the property that every mixed state has a purification, it is necessarily the case that there exists an isomorphism between transformations and bipartite states which has the same structural properties as the  Choi-Jamio\l{}kowski isomorphism. This paper is a good example of what I have described as the study of intratheoretic causal structure, but the structure proposed is  quite different to the one I have suggested here. First, I have taken the `operational Choi-Jamio\l{}kowski isomorphism' to be fundamental and derived other features of quantum mechanics from it, whereas in 
 ref \cite{Chiribella} the operational Choi-Jamio\l{}kowski is derived from other features of the theory, so the intratheoretic causal structure is in fact precisely reversed. Second, I have specifically avoided using the concept of state in my formulation of the operational Choi-Jamio\l{}kowski isomorphism and instead derived properties of the state from the isomorphism together with constraints on possible correlations, whereas in ref \cite{Chiribella} one of the  axioms from which the rest of quantum mehcanics is derived is `every mixed state has a purification,' so in this approach it seems natural to take the state as a fundamental object and not an emergent feature.

 \section{Mathematical Background}

 \section{Preliminaries \label{new}} 
 \subsection{Operational Theories \label{ot}} 
 
 In attempting to  understand some of the special features of quantum mechanics, it is often helpful to make comparisons between quantum theory and other possible theories. In this context it is common to employ the language of generalized probabilistic theories, where theories are defined entirely in terms of preparation procedures and measurements, eschewing abstract mathematical constructions. This has been a very active area of study in recent years, with many interesting results emerging to show how quantum mechanics relates to the broader field of possible operational theories.   For example, it was shown that quantum mechanics is not the maximally non-local theory without signalling\cite{Rohrlich}:  there exists a gap between the numerical bound on the set of non-local correlations which can be produced in a theory limited only by no-signalling, and the corresponding `Tsirelson' bound on the set of non-local correlations which can be produced in quantum theory, and much effort has gone into trying to explain this difference (see for example refs \cite{Niestegge,  CabelloSE, Bub2, Pawlowski}).

 In the framework of operational theories, a given theory is  specified as a quadruple $(\mathcal{P},\mathcal{M}, \mathcal{T}, p)$ where $\mathcal{P}$ is a set of preparations, $\mathcal{M}$ is a set of measurements, $\mathcal{T}$ is a set of transformations, and the function $p(M^x | P, T) $ specifies, for every possible combination of  preparation $P$, transformation $T$, and measurement outcome $M^x$, the probability of obtaining outcome $M^x$ to the measurement  $M$ if it is performed on a system prepared according to $P$ and then subjected to transformation $T$. \cite{spekkenscontextuality}

 I will sometimes write $p(M^x | P) $ as shorthand for  $p(M^x | P, \mathbb{I}) $, so $p(M^x | P) $ denotes the probability of obtaining outcome $M^x$ to the measurement  $M$ if it is performed directly on a system prepared according to $P$. For the operational theories considered in this paper,  I stipulate that performing a  transformation followed by a given measurement is equivalent to simply performing some other measurement, i.e. for any $P, T, \{ M^x\}$, there exists some $\{ N^x\}$ such that $p(M^x | P, T) = p(N^x | P, \mathbb{I})$; for any operational theory where this is not the case, we can trivially make it the case by expanding the set $\mathcal{M}$.

 \subsection{Ensemble preparations} 
 
 The operational formulation of the Choi-Jamio\l{}kowski isomorphism will make use of the following concept:
 
 \begin{definition}
 	An \emph{ensemble preparation}, $P$, specified by a probability distribution $p(i) : i \in   \{ 1, 2 ... N \}$ and a set of preparations $ \{ Q_i : i \in   \{ 1, 2 ... N \} \} $, is a procedure  in which an observer draws a number $i$ from $ \{ 1, 2 ... N \} $ with probability distribution $p(i)$, and then performs the corresponding preparation $Q_i$. 
 \end{definition}

 In particular, when the operational theory in question is quantum mechanics, every possible ensemble preparation can be described by a  POVM $\{ M_i \}$ and density operator $\rho$, where  $p(i) = Tr(\rho M_i)$ and $P_i$ is a preparation which produces the quantum state $\rho_i =  \frac{\sqrt{\rho} M_i \sqrt{\rho} }{Tr(M_i \rho)} $ \cite{Leiferconditional}. 
 
 \vspace{2mm} 
 
 For brevity, I will also use the following definitions:

 \begin{definition} 
 	For any ensemble preparation $P$ for a given system $S$, any set of channels $ T_2,T_3, ... T_n$ which may then   be applied to  $S$, and any set of measurements   $M_2, M_3 ... M_n$ which can be applied to the distinct outputs of these channels, I denote by $p_{P; T_2 ... T_n ;M_2 ... M_n}$  the joint probability distribution over the outcome of the random choice in the ensemble preparation $P$ and the outcome of the measurements $M_2, M_3 ... M_n$.
 	
 \end{definition}
 \begin{figure}
 	\begin{center}
 		
 		\begin{tikzpicture}[scale=0.7]
 		
 		%define the call points, the end points, and the start point

 		\coordinate (Ca) at (0,-3);
 		\coordinate (Cb) at (0,0);
 		\coordinate (Cc) at (2,0);
 		\coordinate (Cd) at (-2,0);
 		
 		%draw the frame
 		\draw[->] (Ca) -- (Cb) ;
 		\draw[->] (Ca) -- (Cc) ;
 		\draw[->] (Ca) -- (Cd) ;

 		\node [below right] at (1,-1.5) {$p(3)$};
 		\node [right] at ( 0, -0.7) {$p(2)$};
 		\node [below left] at (-1, -1.5) {$p(1)$};
 		
 		\node [above] at (2,0 ) {$3$};
 		\node [above] at (Cb) {$2$};
 		\node [above] at (Cd) {$1$};
 		
 		\draw[->] (2, 1) -- (2,2) ;
 		\draw[->] (0,1) -- (0,2) ;
 		\draw[->] (-2,1) -- (-2,2) ;
 		
 		\node [above] at (2,2 ) {$Q_3$};
 		\node [above] at (0, 2) {$Q_2$};
 		\node [above] at (-2,2) {$Q_1$};

 		\draw[->] (2, 3) -- (0.3,6) ;
 		\draw[->] (0,3) -- (0,6) ;
 		\draw[->] (-2,3) -- (-0.3,6) ;
 		
 		\draw[->] (0,6.2) -- (0,7.2);
 		
 		\node [above] at (0,7.2) {$M$};
 		
 		\node [right] at (0,6.5) {$T$};
 		
 		\draw[->] (0, 8) -- (0,11) ;
 		\draw[->] (0,8) -- (2,11) ;
 		\draw[->] (0,8) -- (-2,11) ;

 		\node [above] at (0,11) {$M^2$};
 		\node [above] at ( 2,11) {$M^3$};
 		\node [above] at (-2,11) {$M^1$};

 		\end{tikzpicture}
 		
 		\caption{A schematic diagram of an ensemble preparation $P$ with three possible results for the random number generation step, followed by a transformation $T$, followed by a  measurement $M$ with three possible outcomes $M^1, M^2, M^3$. This  scenario can be described by the probability distribution $p_{P; T ;M }$. }
 		
 		\label{prep}
 	\end{center}
 \end{figure}
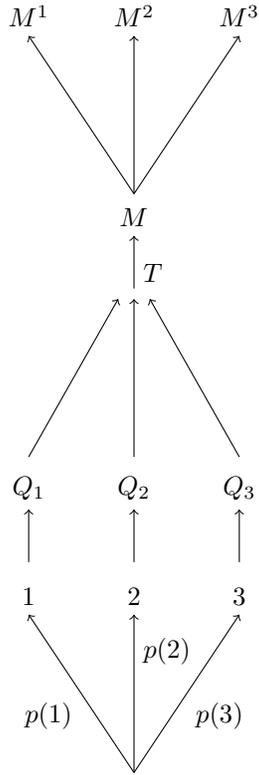

 \begin{definition}

 	For any joint preparation $P_{12 ... n}$ for a set of systems $S_1, S_2 ... S_n$, and any set   of measurements $M_1, M_2, ... M_n $ which can be applied separately to systems $S_1, S_2, ... M_n $, I denote by $p_{P_{12... n}; M_1 ... M_n}$ the  joint probability distribution over the outcomes of the measurements $M_1, M_2, ... M_n $.

 \end{definition}

 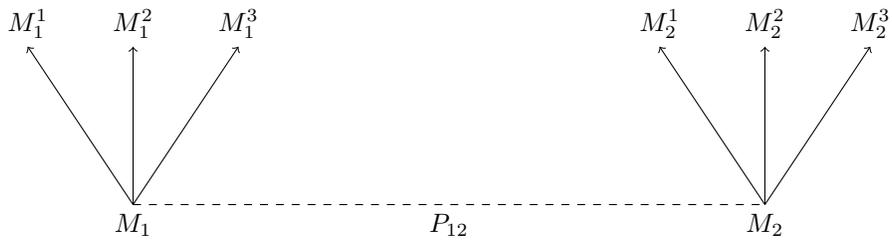
\begin{figure}
 	\begin{center}
 		
 		\begin{tikzpicture}[scale=0.7]
 		
 		%define the call points, the end points, and the start point

 		%draw the frame
 		\draw[->] (-6, 0) -- (-4,3) ;
 		\draw[->] (-6,0) -- (-6,3) ;
 		\draw[->] (-6,0) -- (-8,3) ;
 		
 		\draw[dashed] (-6,0) -- (6,0);
 		
 		\node [below] at (6,0) {$M_2$};
 		
 		\node [below] at (-6,0) {$M_1$};
 		
 		\node [below] at (0,0) {$P_{12}$};
 		
 		\draw[->] (6, 0) -- (4,3) ;
 		\draw[->] (6,0) -- (6,3) ;
 		\draw[->] (6,0) -- (8,3) ;
 		
 		\node [above] at (4,3) {$M_2^1$};
 		\node [above] at (6,3) {$M_2^2$};
 		\node [above] at (8,3) {$M_2^3$};
 		\node [above] at (-8,3) {$M_1^1$};
 		\node [above] at (-6,3) {$M_1^2$};
 		\node [above] at (-4,3) {$M_1^3$};

 		\end{tikzpicture}
 		\caption{A schematic diagram of a joint preparation $P_{12}$ on a bipartite state, followed by  followed by measurement $M_1$ with possible outcomes $M_1^1, M_1^2, M_1^3$ on one system, and measurement $M_2$ with possible outcomes $M_2^1, M_2^2, M_2^3$ on the other system.  This  scenario can be described by the probability distribution $p_{P_{12}; M_1, M_2 }$. }
 		
 		\label{bipartite}
 	\end{center}
 \end{figure}

 \subsection{The Choi-Jamio\l{}kowski Isomorphism }

 In its original form, the Choi-Jamio\l{}kowski isomorphism is a one-to-one map between unnormalized quantum states and completely positive (CP) operators. Specifically, for a given CP operator $\epsilon_{B|A}$ mapping a Hilbert space $\mathscr{H}_A$ to a Hilbert space $\mathscr{H}_B$, the corresponding Choi-Jamio\l{}kowski state $ \rho^{\epsilon}  $ is obtained by applying $\epsilon_{B|A} $ to half of the maximally entangled state $ \rho_{A A'} =  \frac{1}{d} \sum_i \sum_{i'} | i \rangle_A |  i \rangle_{A'}  \langle  i' |_A  \langle    i' |_{A'}  $, where $A'$ is a quantum system with the same Hilbert space as $A$, and $d$ is the dimension of this Hilbert space:	\cite{Leiferconditional, Bengtsson}\footnote{The formula presented here is for the Choi version of the isomorphism; the Jamio\l{}kowski version is $\rho ^{\epsilon}_{AB}   = ( \epsilon_{B | A} \otimes \mathbb{I} ) (  \frac{1}{d} \sum_i \sum_{i'} | i \rangle_A |  i \rangle_{A'}  \langle  i' |_A   \langle  i'  |_{A'}  )^{T_A})$, where $T_A$ refers to the partial transpose in the basis  $\{ | i \rangle \}$. The key difference between the two formulations is that the  Jamio\l{}kowski version, unlike the Choi version, is basis-independent. But since I will not henceforth be working within the quantum formalism, this distinction is not important.}
 
 \[ \rho^{\epsilon}_{AB}   = ( \epsilon_{B | A} \otimes \mathbb{I} ) \rho_{A A'}  \]

 However, in ref \cite{Leiferconditional}  the isomorphism is reformulated as a map between conditional density operators and completely positive trace-preserving  (CPTP) operators. This formulation, which I paraphrase below, is more transparent in its physical interpretation and hence I will take it as the starting point for the operational approach:

 \begin{lemma} 
 	For any  bipartite state  $\rho^{\epsilon}_{AB}$, there exists a CPTP map $\epsilon$ and a reduced state $\rho_A = Tr_B(\rho^{\epsilon}_{AB})$ such that given any two POVMs $M$ and $O$, if $M^T$ is obtained by taking the transpose of all the measurement operators in $M$ with respect to some fixed basis, then when we perform the ensemble preparation described by the POVM $M^T$ and the density operator $\rho_A$, then evolve the state according to $\epsilon$, then perform the measurement $O$, the probability that state $\rho_i$ is prepared and then the measurement $M$ has outcome $j$  is the same as the joint probability of obtaining outcomes $M_i$ and $O_j$ when the POVM $M$ is performed on system $A$ and the POVM $O$ is performed on system $B$ for a bipartite system $AB$ in the state $\rho^{\epsilon}_{AB}$. 
 	
 	Conversely, for any pair of a CPTP map and state $\rho$ there exists a bipartite state $\rho^{\epsilon}_{AB}$ such that the same conditions hold, so we have defined an isomorphism between bipartite states and pairs $(\rho_A, \epsilon^r)$, where $ \epsilon^r$ denotes the restriction of the CPTP map $\epsilon$ to the support of $\rho_A$.

 	\label{CJlemma} 
 	
 \end{lemma}

 \subsection{Reformulation} 
 
 Using these concepts, we can define an operational version of the Choi--Jamio\l{}kowski isomorphism.  
 \begin{definition} 
 	
 	\textbf{Operational Choi-Jamio\l{}kowski Isomorphism:}

 	\vspace{2mm} 
 	For any  joint preparation  $P_{1 2 3 ... n}$ on a set of systems $S, S_2, ... S_n$, there exists a set of channels $ T_2,T_3, ... T_n$ which may simultaneously be applied to the system $S$,  such that  for any set of measurements $M$, $M_2$, ... $M_n$ which may  be performed on $S, S_2, ... S_n$,  there exists an ensemble preparation $P$ for the system $S$ such that  	the distribution $p_{P_{123 ... n} ; M, M_2 ... M_n}$  is the same as the distribution $ p_{P ; T_2  ...  T_n ;M_2 ... M_n}$.  
 	
 	\vspace{2mm} 
 	Conversely, for any set of channels $T_2, T_3, ... T_n$ which may simultaneously be applied to the system $S$ to produce a set of systems  $ S_2, ... S_n$,  there exists a 	joint preparation  $P_{1 23 ... n}$  for systems $S, S_2, S_3 ... S_n$ such that  for any ensemble preparation $P$ which may be performed for system $S$ and any set of measurements  $M_2, M_3, ... M_n$ which may  be performed on the products $ S_2, ... S_n$, there exists a measurement $M$ on $S$ such that  	the distribution $p_{P_{123 ... n} ; M  M_2 ... M_n}$  is the same as the distribution $ p_{P ; T_2  ...  T_n ; M_2, ... M_n}$.

 	\label{ontological}

 \end{definition} 
 
\begin{figure}
 	\begin{center}
 		\begin{tikzpicture}[scale=0.7]
 		
 		%define the call points, the end points, and the start point

 		%draw the frame
 		\draw[->] (-9, 0) -- (-7,3) ;
 		\draw[->] (-9,0) -- (-9,3) ;
 		\draw[->] (-9,0) -- (-11,3) ;
 		
 		\draw[dashed] (-9,0) -- (9,0);
 		
 		\node[below] at (0,-1) {$P_{123}$};
 		\node[below] at (-9,0) {$M_1$};
 		
 		\node[below] at (9,0) {$M_3$};
 		
 		\node[below] at (0,0) {$M_2$};
 		
 		\node[above] at (-11,3) {$M_1^1$};
 		\node[above] at (-9,3) {$M_1^2$};
 		\node[above] at (-7,3) {$M_1^3$};

 		\draw[->] (9, 0) -- (7,3) ;
 		\draw[->] (9,0) -- (9,3) ;
 		\draw[->] (9,0) -- (11,3) ;
 		
 		\node[above] at (7,3) {$M_3^1$};
 		\node[above] at (9,3) {$M_3^2$};
 		\node[above] at  (11,3) {$M_3^3$};
 		
 		\draw[->] (0, 0) -- (-2,3) ;
 		\draw[->] (0,0) -- (0,3) ;
 		\draw[->] (0,0) -- (2,3) ;
 		
 		\node[above] at (-2,3) {$M_2^1$};
 		\node[above] at  (0,3) {$M_2^2$};
 		\node[above] at (2,3) {$M_2^3$};

 		\end{tikzpicture}
 		
 		\caption{A schematic diagram of a joint preparation $P_{12}$ on a bipartite state, followed by  followed by a measurement $M_1$ with possible outcomes $M_1^1, M_1^2, M_1^3$ on one system, and measurement $M_2$ with possible outcomes $M_2^1, M_2^2, M_2^3$ on the other system.  This  scenario can be described by the probability distribution $p_{P_{123}; M_1, M_2, M_3 }$. }
 		
 		\label{tripartite}
 		
 	\end{center}
 \end{figure}
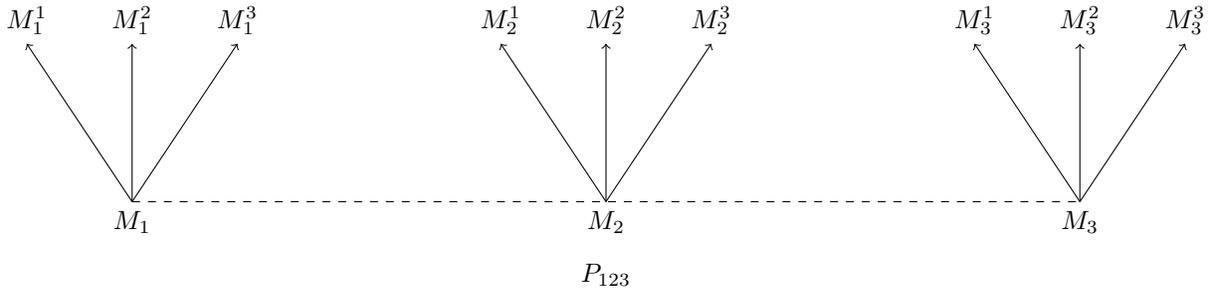

 In graphical terms, this says that for any scenario as in figure \ref{prep} there exists a scenario as in figure \ref{bipartite} and vice versa; likewise, for any scenario as in figure \ref{prep2} there exists a scenario as in figure \ref{tripartite} and vice versa. 
 
 \begin{figure}
 	\begin{center}
 		
 		\begin{tikzpicture}[scale=0.7]
 		
 		%define the call points, the end points, and the start point

 		\coordinate (Ca) at (0,-3);
 		\coordinate (Cb) at (0,0);
 		\coordinate (Cc) at (2,0);
 		\coordinate (Cd) at (-2,0);
 		
 		%draw the frame
 		\draw[->] (Ca) -- (Cb) ;
 		\draw[->] (Ca) -- (Cc) ;
 		\draw[->] (Ca) -- (Cd) ;

 		\node [below right] at (1,-1.5) {$p(3)$};
 		\node [right] at ( 0, -0.7) {$p(2)$};
 		\node [below left] at (-1, -1.5) {$p(1)$};
 		
 		\node [above] at (2,0 ) {$3$};
 		\node [above] at (Cb) {$2$};
 		\node [above] at (Cd) {$1$};
 		
 		\draw[->] (2, 1) -- (2,2) ;
 		\draw[->] (0,1) -- (0,2) ;
 		\draw[->] (-2,1) -- (-2,2) ;
 		
 		\node [above] at (2,2 ) {$Q_3$};
 		\node [above] at (0, 2) {$Q_2$};
 		\node [above] at (-2,2) {$Q_1$};

 		\draw[->] (2, 3) -- (0.3,6) ;
 		\draw[->] (0,3) -- (0,6) ;
 		\draw[->] (-2,3) -- (-0.3,6) ;
 		
 		\draw[->] (0,6) -- (-4,8);
 		\draw[->] (0,6) -- (4,8);
 		
 		\node [below left] at (-2,7) {$T_2$};
 		
 		\node [below right] at (2,7) {$T_3$};
 		
 		\node [below] at (-4,8) {$M_2$};
 		\node [below] at (4,8) {$M_3$};

 		\draw[->] (-4, 8) -- (-4,11) ;
 		\draw[->] (-4,8) -- (-2,11) ;
 		\draw[->] (-4,8) -- (-6,11) ;

 		\node [above] at (-2,11) {$M_2^3$};
 		\node [above] at (-4,11) {$M_2^2$};
 		
 		\node [above] at (-6,11) {$M_2^1$};
 		
 		\draw[->] (4, 8) -- (4,11) ;
 		\draw[->] (4,8) -- (2,11) ;
 		\draw[->] (4,8) -- (6,11) ;

 		\node [above] at (2,11) {$M_3^1$};
 		\node [above] at (4,11) {$M_3^2$};
 		
 		\node [above] at (6,11) {$M_3^3$};

 		\end{tikzpicture}
 		
 		\caption{A schematic diagram of an ensemble $P$ with three possible results for the random number generation step, followed by two channels $T2$ and $T_3$ each producing a distinct system, followed by a measurement $M_2$ with three possible outcomes $M_2^1, M_2^2, M_2^3$ on one system, and a measurement $M_3$ with three possible outcomes $M_3^1, M_3^2, M_3^3$ on the other system. This  scenario can be described by the probability distribution $p_{P; T_2, T_3 ;M_2, M_3 }$. }
 		
 		\label{prep2}
 	\end{center}
 \end{figure}
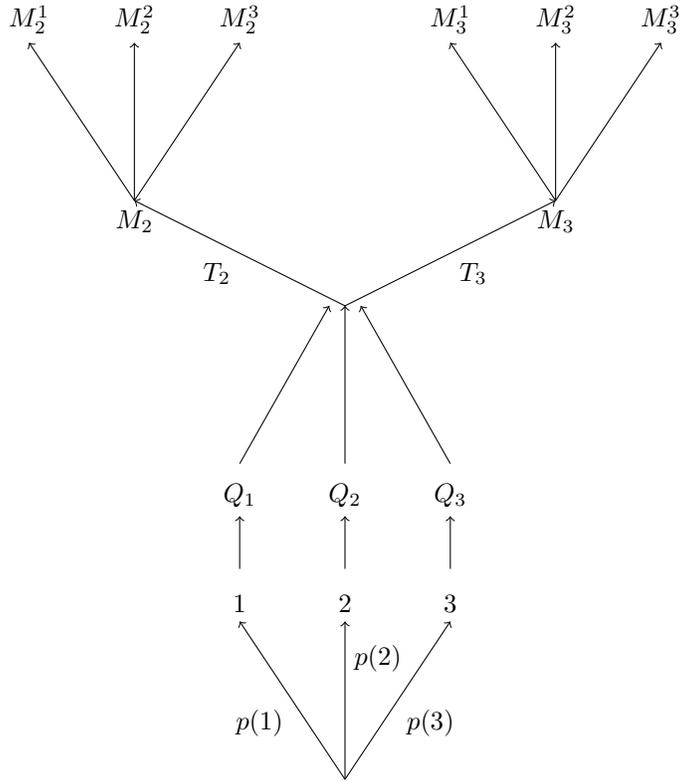

 \subsubsection{Relation to original Choi-Jamio\l{}kowski isomorphism} 
 
 This operational definition of the isomorphism differs from the original in several respects. It is of course to be expected that an operational formulation of the isomorphism will not be exactly equivalent to the original quantum-mechanical statement of it, since the quantum-mechanical version is heavily theory-dependent and it is not possible to jettison these theory-dependent elements without losing some content. I believe nonetheless that this formulation captures much  of what is important about the original Choi-Jamio\l{}kowski isomorphism.
 
 First, it should be noted that the operational formulation does not actually postulate the existence of an isomorphism: it makes a statement of the form `for each x there exists a y, and for each y there exists an x,' but this does not imply that there is an isomorphism between the sets $\{ x\}$, $\{ y \}$, since I do not insist that each distinct $x$ should be mapped to a different $y$. One could of course make this further stipulation and it might turn out that further interesting results would arise from the correspondingly stronger constraint, but it was not necessary to do this for any of the results obtained here and I thought it best to employ the weakest possible version of the constraint that was still sufficient to derive these results. I have continued to use the term `isomorphism' for consistency with the established terminology.

 Second, the operational formulation applies to all multipartite states, whereas the quantum-mechanical Choi-Jamio\l{}kowski isomorphism applies only to bipartite states. This change has been made in order to make it possible to derive the no-cloning result in a particularly simple and transparent way. This particular generalization has in fact been considered before: it was used to demonstrate the connection between no-cloning/no-broadcasting theorems and the monogamy of entanglement in ref \cite{Leifer22}.  Furthermore, recall that the motivation for putting forward this operational  Choi-Jamio\l{}kowski isomorphism is the intuition that the quantum isomorphism expresses a deep fact about the underlying ontology of the theory - channels behave like preparations because they are in fact the same sort of thing at the level of the underlying ontology. If this is the case, then we would naturally expect the constraint to apply to all multipartite states rather than just bipartite ones, since bipartite states are simply a special case of multipartite states.

 \section{No-broadcasting theorem and the monogamy of entanglement \label{no-broadcasting}}

 \subsection{Background} 
 
 It is a particularly distinctive feature of quantum information that quantum states cannot be cloned - that is, there is no quantum operation which makes perfect copies of two non-orthogonal quantum states. In fact, the impossibility of cloning quantum states is a special case of the impossibility of broadcasting quantum states: 
 
 \begin{definition} 
 	
 	A map $M$ from the Hilbert space $\mathscr{H}$ to the the Hilbert space $\mathscr{H}_A \otimes \mathscr{H}_B$ \textbf{broadcasts} the set of states $S$ iff for any state $\rho \in S$, $Tr_A(M(\rho)) = \rho$ and $Tr_B(M(\rho)) = \rho$. 
 \end{definition} 
 
 \begin{definition} A \textbf{universal broadcasting map} broadcasts any set of states. 
 	
 \end{definition}

 \begin{theorem} \textbf{No-Broadcasting Theorem:} A set of quantum states can be broadcasted if and only if they commute pairwise \cite{BarnumJozsaetc}\footnote{A pair of quantum states $\rho_0, \rho_1$ are said to commute iff $\rho_0 \rho_1 - \rho_1 \rho_2 = 0$\cite{BarnumJozsaetc}}
 	
 \end{theorem} 
 
 \begin{corollary} 
 	There is no universal broadcasting map in quantum mechanics 
 	\label{corollary} 
 \end{corollary}

 \subsection{Operational Formulations \label{correspondencebroadcasting}} 
 
 In  ref \cite{Leifer22}, it is shown that corollary \ref{corollary} can be derived directly from the monogamy of entanglement using the Choi-Jamio\l{}kowski isomorphism: 
 
 \begin{theorem} 
 	
 	Supposing the existence of a universal broadcasting map is equivalent to supposing the existence of a tripartite state of a system $A, B, C$ where both the bipartite reduced state of $A, B$ and the bipartite reduced state of $B, C$ are pure and maximally entangled.
 	\footnote{Ref \cite{Leifer22} also derives the no-broadcasting theorem in full generality from the monogamy of entanglement, but I have not pursued that result here, because the proof relies on facts about the fixed point set of TPCP maps, and 
 		in the operational framework facts concerning the detailed structure of quantum operations are not available to us. }
 	
 \end{theorem}
 
 Here I will derive a similar result, but my derivation will be presented in the context of  general operational theories without presupposing the Hilbert space structure of quantum mechanics. Therefore before embarking on the proof, I need to provide definitions of broadcasting and monogamy which are suited to the operational context.

 In order to define `broadcasting' we will use the concept of as an `operational state';  note, again, that this concept will be used only to define broadcasting, and will not appear in the definitions of the  features from which no-broadcasting will be derived. 
 
 \begin{definition}
 	Given two preparation procedures $P_a, P_b$ which appear in an operational theory, I say that these procedures produce the same \emph{operational state} iff  
 	when a system is prepared using one of these procedures, there is no subsequent measurement or sequence of measurements which can give us any information about whether the system was prepared using $P_a$ or $P_b$. 
 	
 \end{definition}

 \begin{definition} 
 	
 	A transformation $T$ \textbf{broadcasts} the set of operational states $S$ iff when $T$ is applied to a system whose operational state is $\rho \in S$, together with one or more ancilla systems, the output includes two distinct systems both having the operational state $\rho$.
 	
 \end{definition} 

 \begin{definition} 
 	A \textbf{universal broadcasting map} broadcasts any set of operational states.

 \end{definition} 
 
 Defining the relevant monogamy property for an operational context is less straightforward. In quantum mechanics, the monogamy of entanglement refers to the fact that the amount of entanglement a quantum system has with a given system limits the amount of entanglement it can have with any other system - in the most extreme case, a quantum system which is maximally entangled with a given quantum system cannot be entangled at all with any other quantum system. But entanglement is a property of quantum states, usually quantified using state-dependent measures like the `concurrence' (or `tangle')\cite{Koffmanetal}, and these measures do not have a   straightforward physical interpretation, making it difficult to translate them to the framework of operational theories. 
 
 Therefore it is more helpful for us to   focus on  the `monogamy of correlations,' which is a property of certain theories in which there are limits on how strongly a system can be correlated with other systems. The monogamy of correlations is defined purely in terms of observable statistics, and is therefore much more easily cast in an operational framework. 
 
 In order to facilitate comparison to the quantum case, I will define monogamy for a general operational theory using the CHSH quantity:
 
 \begin{definition} 
 	\textbf{CHSH quantity:}
 	For any joint preparation $P_{AB}$ of two systems, and any pair of measurements $M_A^0, M_A^1$ which may be performed on system $A$, and any pair of measurements $M_B^0, M_B^1$ which may be performed on system $B$, the CHSH quantity $  \mathscr{B}_{AB}(  P_{AB}, M_A^0, M_A^1, M_B^0, M_B^1) $ for this combination of preparation and measurements  	 is defined by the following sum of expectation values: 
 	
 	\begin{multline}  \mathscr{B}_{AB}(  P_{AB}, M_A^0, M_A^1, M_B^0, M_B^1)  : = \\ \hspace{-10mm} \langle A B \rangle_{P_{AB}, M_A^0, M_B^0}  + \langle A B \rangle_{P_{AB}, M_A^0, M_B^1} +\langle  A B \rangle_{P_{AB}, M_A^1, M_B^0}- \langle A B\rangle_{P_{AB}, M_A^1, M_B^1}  \\ \label{Belloperator} \end{multline}

 \end{definition} 
 
 Note that is also possible to define a related CHSH quantity $ \mathscr{B}_{AB}(  P^0, P^1, T,   M_B^0, M_B^1) $ for the scenario where an experimenter applies one of two ensemble preparations $P^0, P^1$ to a   system, then applies a fixed transformation  $T$,   then performs one of two measurements  $M_B^0, M_B^1$. This quantity is defined  just as in equation \ref{Belloperator}, where  the result of the measurement $M_A$ is replaced with the result of the probabilistic choice in the ensemble preparation. This definition ensures that  if for some transformation $T$, some set of preparations  $P_{AB}$, $P^0, P^1$ and some set of  measurements $M^0_1, M^1_1$,   $M^0_2, M^1_2$, for any $x, y \in \{0, 1\}$ the distributions  	$p_{P_{AB} ; M^x_1 M^y_2}$ and $p_{P^x ;T ;M^y_2}$ are the same, then $ \mathscr{B}_{AB}(  P_{AB}, M_A^0, M_A^1, M_B^0, M_B^1) =  \mathscr{B}_{AB}(  P^0, P^1, T,   M_B^0, M_B^1)$. 
 
 With a definition for the CHSH quantity in hand, we may now define two types of  monogamy which might be obeyed by an operational theory:
 
 \begin{definition} 
 	\textbf{Non-signalling monogamy of correlations:} An operational theory obeys no-signalling monogamy of correlations iff for any joint preparation of three systems $S_A, S_B, S_C$, for any choice of measurements $M_A^0, M_A^1$ on $S_A$, any choice of measurements $ M_B^0, M_B^1$ on $S_B$, and any choice of measurements $M_C^0, M_C^1$ on $S_C$, the associated CHSH quantities satisfy:
 	
 	\[ \mathscr{B}_{AB}(P_{ABC}, M_A^0, M_A^1, M_B^0, M_B^1)  + \mathscr{B}_{BC}(P_{ABC}, M_B^0, M_B^1, M_C^0, M_C^1)  \leq 4 \]  \label{monogamyq} \end{definition}
 
 It can be shown that this bound is necessarily obeyed by any  operational theory which does not allow signalling (see section \label{signalling})\cite{Toner}.

 \begin{definition} 
 	\textbf{Strong monogamy of correlations:} An operational theory obeys strong monogamy of correlations iff for any joint preparation of three systems $S_A, S_B, S_C$, for any choice of measurements $M_A^0, M_A^1$ on $S_A$, any choice of measurements $ M_B^0, M_B^1$ on $S_B$, and any choice of measurements $M_C^0, M_C^1$ on $S_C$, the associated CHSH quantities satisfy:
 	
 	\[ \mathscr{B}_{AB}(P_{ABC}, M_A^0, M_A^1, M_B^0, M_B^1)^2 + \mathscr{B}_{BC}(P_{ABC}, M_B^0, M_B^1, M_C^0, M_C^1)^2 \leq 8 \]  \label{monogamyq} \end{definition}
 
 Strong monogamy of correlations need not be obeyed by all non-signalling operational theories, but it \emph{is} obeyed by quantum mechanics\cite{Toner2}. 
 
 \vspace{2mm}

 \subsection{Bell nonlocality}
 We will also require the notion of Bell nonlocality, which is again defined by analogy with the quantum case:

 \begin{definition}

 	An operational theory exhibits Bell nonlocality iff there exists at least one joint preparation  $P_{AB}$ for two systems $S_A$ and $S_B$, such that for some pair of measurements  $M_A^0, M_A^1$ on $S_A$ and some pair of measurements $ M_B^0, M_B^1$ on $S_B$,   the CHSH quantity    satisfies:  \[ \mathscr{B}_{AB}(P_{AB}, M_A^0, M_A^1, M_B^0, M_B^1) > 2 \]

 \end{definition} 
 
 \subsection{theorem \label{deriv}}
 Using these concepts, we have the following theorem (see appendix \ref{app1} for the proof). 
 \begin{theorem} \label{theorem1}
 	
 	In an operational theory which obeys the operational Choi-Jamiolkowksi isomorphism and exhibits Bell nonlocality,  the existence of a universal broadcasting map implies that the theory violates strong monogamy of correlations
 	
 \end{theorem}
 
 Note that if we impose the ontological Choi-Jamio\l{}kowski isomorphism as a fundamental constraint on an operational theory, it follows by modus tollens that the strong monogamy of correlations together with the existence of Bell nonlocality  implies the absence of broadcasting, as advertised in table \ref{corr}.

 \subsection{Quantum interference}
 
 The results of section \ref{deriv} may also be used to derive the existence of quantum interference from the monogamy of correlations together with the existence of Bell non-locality. 
 
 To see this,  let us limit ourselves to considering GPTs whose observables have the algebraic
 structure of the self-adjoint operators in a $C^*$ algebra. Not all GPTs meet this criterion\footnote{See ref \cite{Halvorson_2004} for a discussion of the general conditions under which a theory can be represented as a $C^*$ algebra}, but nonetheless it is a reasonable heuristic framework since a large class of important GPTs, including classical mechanics and quantum mechanics, can indeed be thus represented. Then observe that for any theory represented as a $C^*$-algebra, it can be shown that if the $C^*$ algebra is commutative then the theory has a univeral broadcasting map \cite{Clifton_2003, Bub}. Thus if a GPT does not have a universal broadcasting map, any    $C^*$ algebra that represents it must be non-commutative. Finally, note that theories represented by commutative $C^*$ algebras are classical phase space theories, whereas theories represented by non-commutative $C^*$ algebras have observables which cannot be simultaneously measured and hence exhibit superposition and interference\cite{Clifton_2003}. 
 
 Thus, if we limit ourselves to GPTs which have the algebraic
 structure of the self-adjoint operators in a $C^*$ algebra and which obey the operational Choi-Jamio\l{}kowski isomorphism, it follows from theorem \ref{theorem1} that any GPT which exhibits strong monogamy of correlations and Bell non-locality must also exhibit superposition and interference. This means that superposition and interference need not be regarded as properties of the quantum state - with the help of the operational Choi-Jamio\l{}kowski isomorphism they can be derived directly from properties of non-local correlations, with no need to appeal to a concept of state at all.

 \section{Preparation Contextuality and No-Signalling \label{contextualitysec}}

 \subsection{Background} 
 
 One of the most puzzling feature of quantum theory is the fact that it is \emph{contextual}: that is, it is not possible to interpret the outcomes of quantum-mechanical measurements as if they simply reveal  the pre-existing value of some property of the system being measured. 
 
 The original formulation of contextuality\cite{sep-kochen-specker} tied the concept of contextuality to the concept of determinism, but the notion was subsequently generalized by Spekkens\cite{Spekkens}. Spekkens' form of contextuality is most easily defined within the ontological models framework, where one supposes that every system has a single real `ontic state,' which determines the probabilities for the outcomes of any measurement on that system\footnote{The use of this faramework does not necessarily imply the endorsement of a description of quantum reality in terms of underlying states; the ontological models approach could simply be regarded as a helpful language in which I may express   mathematical facts about the structure of quantum theory, such as its contextual character.}. An ontological model thus consists of a space $\Lambda$ of ontic states $\lambda$, a set of probability distributions $\mu^P(\lambda)$ giving the probability that the system ends up in the state $\lambda$ when we perform the preparation procedure $P$, a set of response functions $\xi_{M,X}(\lambda)$ giving the probability that we obtain outcome $M^x$ when we perform measurement $M$ on a system whose ontic state is $\lambda$, and a set of column-stochastic matrices $T^O$ representing the way in which the ontic state is transformed when some operation $O$ is applied to the system. 	A valid ontological model must satisfy the following positivity and normalization conditions: 
 
 \begin{enumerate} 
 	
 	\item $\forall P, \lambda \quad \mu_P(\lambda) \in [ 0, 1]$
 	
 	\item $\forall P \quad \sum_{\lambda} \mu_P(\lambda) = 1$
 	
 	\item $\forall M, X \quad \vec{\xi}_{M,X}(\lambda) \in [0, 1]$
 	
 	\item $\forall M  \quad \sum_X \vec{\xi}_{M,X}(\lambda) = 1$
 	
 	\item $\forall O \quad T_0$ is a column-stochastic matrix. \footnote{A column-stochastic matrix is a matrix whose entries are all non-negative and whose columns sum to one. Left-multiplication by a column-stochastic matrix is the most general possibility for the representation of  operations in an ontological model, since such an operation must map ontic state $\lambda_i$ to $\lambda_j$ with some probability, which is specified by entry $(i, j)$ in the transformation matrix. Given that the entries are all probabilities, they must be nonnegative, and given that each transformation must map $\lambda_i$ to some state, the sum of the entries in a column must be one.}

 \end{enumerate}

 An ontological model is \emph{preparation contextual} if it does \emph{not} represent every  quantum state by a unique probability distribution $\mu(\lambda)$; \emph{transformation contextual} iff it does \emph{not} represent every possible quantum operation  $O$ by  a unique transformation matrix $T_O$; and \emph{measurement contextual} if it does \emph{not} represent every possible quantum measurement  element $M^x$ by a unique set of response functions $\xi_{M,X}(\lambda)$.	Spekkens proved\cite{Spekkens} that any ontological model of quantum mechanics which reproduces all the correct measurement statistics must exhibit preparation contextuality, but examples such as the Kochen-Specker model \cite{HarriganSpekkens} show that this is not true for measurement contextuality. \footnote{Transformation contextuality has not been studied so thoroughly, and most existing explicit ontological models omit transformations, so to my knowledge the question of whether or not quantum mechanics necessarily exhibits transformation contextuality remains unresolved.}
 
 \subsection{Operational Formulations} 
 
 I now generalize the definition of preparation contextuality to the framework of operational theories. (Similar generalizations can be made for measurement and transformation contextuality, but I will not need these concepts here).

 \begin{definition} 
 	
 	We say that an operational theory is \emph{preparation contextual} iff it is not possible to represent the theory by a valid ontological model in which every operational state is represented by a unique probability distribution over ontic states.

 \end{definition} 
 
 I will also employ a definition of the no-signalling principle suited for an operational framework. This generalization is a straightforward one, since it is just the operational version of the no-signalling principle as it is used in the context of relativistic cryptography\cite{bhk}:

 \begin{definition} \label{signalling}
 	\textbf{Operational no-signalling principle:}  In a  process involving a set of non-communicating devices $\{D_i \} : i \in  \{ 1 \ldots N \}$  such that  device $D_i$ accepts an input variable $N_i$ and produces an output variable  $O_i$, let $J$ be any subset of $ \{ 1 \ldots N \}$, let $O_J$ be the set of variables $\{ O_j : j \in J \}$, let $N_J$ be the set of variables $\{ N_j : j \in J \}$; then if the inputs $\{ N_i \}$ are uncorrelated, the outcomes satisfy $p(O_J | N_1 , \ldots , N_n,  ) = p( O_J | N_J)$. 
 	
 	\label{cryptographic}
 \end{definition}

 \subsection{theorem}
 
 Thus we have the following theorem (see appendix \ref{app1} for the proof).

 \begin{theorem} \label{theorem2}
 	
 	Given an operational theory which  obeys the operational Choi-Jamio\l{}kowski isomorphism and the no-signalling principle, if the theory is preparation non-contextual, it does not exhibit Bell nonlocality. 
 	
 \end{theorem} 
 
 Note that if we impose the ontological Choi-Jamio\l{}kowski isomorphism as a fundamental constraint on an operational theory, it follows by modus tollens that if the theory exhibits Bell nonlocality but obeys the no-signalling principle, it must be preparation contextual;  that is, preparation contextuality is a necessary feature of any such operational theory describing a world which is non-local but also non-signalling.

 \section{Uncertainty Relations \label{uncertainty}}
 
 \subsection{Background}
 
 'Heisenberg's Uncertainty Principle' is one of the most well-known features of quantum physics, not least because the fact that there exist incompatible measurements whose results cannot be simultaneously predicted with certainty seems strongly in conflict with our classical intuitions. This provides a clear imperative to seek an explanation for the existence of such apparently arbitrary limitations on the properties that quantum systems can possess. 
 
 To be more precise, there are in fact a number of different  uncertainty relations  which characterize tradeoffs between the precision with which the outcomes of various sets of measurements can be predicted. Most of these relations were originally formulated in terms of standard deviations  and commutators, but this approach has been criticized on the grounds that the use of commutators makes the relations state-dependent, which should not be the case given that in quantum mechanics there always exists a dynamical evolution which transforms any initial pure state into any any other pure state\cite{Deutschentropic}. Thus more recently there has been a move towards expressing uncertainty relations as a constraint on the Shannon entropies corresponding to the probability distributions over outcomes associated with various possible measurements\cite{Deutschentropic,WehnerWinter,Guo}. However these entropic uncertainty relations   have their own limitations -  in particular, they are not capable of distinguishing between different possible combinations of outcomes\cite{OppenheimWehner} - and therefore   in certain cases it is more useful to consider \emph{fine-grained} uncertainty relations, which directly constrain particular sets of outcomes of different measurements. 
 
 The derivation of this section is inspired by the work of ref \cite{OppenheimWehner}, where it is shown that it is possible to derive the Tsirelson bound from the following    fine-grained entropic uncertainty relation, which applies to any pair of orthogonal two-outcome measurements $M$ and $M'$ on a two-dimensional quantum system   (such   as the  binary spin-half observables $X$ and $Z$):
 
 \begin{equation} \forall m \forall n \forall \rho \in \{ 0, 1 \} \;  p(  m | M)_{\rho} + p(  n | M')_{\rho} \leq 1 + \frac{1}{\sqrt{2}}   \label{finegrained} \end{equation}
 
 Here, $p(  m | M)_{\rho}$ denotes the probability that I obtain outcome $m \in \{ 0, 1\}$ when I perform measurement $M$ on a system in quantum state $\rho$.
 
 \vspace{2mm} 
 
 In the operational framework,
 equation \ref{finegrained} becomes a constraint applying to all preparations in the operational theory which prepare systems of dimension two, according to the operational definition of `dimension' which I provide in section \ref{definitionsuncertainty}. I also reverse the argument of   ref \cite{OppenheimWehner}, in order to derive the uncertainty relation from the Tsirelson bound. I  then note that it was shown in ref \cite{Pawlowski} that the Tsirelson bound can be derived from the principle of information causality: 
 
 \begin{definition} 
 	
 	\textbf{Information causality:} if Alice and Bob pre-share a set of devices which exhibit nonlocal correlations, and Alice  receives a bit string $N_0 N_1 ... N_n$ and sends Bob a classical message $M$ of $m$ bits, and Bob performs a measurement with some setting $k$ and obtains outcome $O$, then $\sum_r I( M O :  N_r  | k = r) > m$ \cite{Pawlowski}  
 	\label{Pawlowski} 
 	
 \end{definition} 
 
 Thus, if we take it that information causality is  a fundamental constraint on quantum theory, as suggested in \cite{Pawlowski}, the fact that quantum states obey this particular fine-grained uncertainty relation in quantum theory may be understood as a consequence of the fundamental constraint. Moreover, in quantum mechanics, the inequality in equation \ref{finegrained} is tight (for example, the inequality is saturated if we choose measurements $X$, $Z$ and prepare the system to be measured in an eigenstate of $\frac{1}{\sqrt{2}} (X + Z)$ or 
 $\frac{1}{\sqrt{2}} (X - Z)$) and thus  argue in this particular case information causality may be understood as setting the exact limits of quantum uncertainty.

 \subsection{Operational Formulations \label{definitionsuncertainty}} 
 
 \vspace{2mm} 
 
 Since the fine-grained uncertainty relation of equation \ref{finegrained} applies only for systems of dimension two, we need an operational way to single out systems of dimension two. To do so, we must first add some structure to the usual framework of operational theories. Thus far, we have assumed that for a given operational theory $(\mathcal{P},\mathcal{M}, \mathcal{T}, p)$ it is possible to perform any preparation followed by any transformation followed by any measurement, but in a theory with a meaningful concept of dimension this is not so, because we can only perform measurements which match the dimension of the system I have prepared. Thus in this section I consider an operational theory to be a collection of \textbf{subtheories} $(\mathcal{P},\mathcal{M}, \mathcal{T}, p)$, such that within any given subtheory it is  possible to perform any preparation followed by any transformation followed by any measurement, but it is not possible to perform a preparation from one subtheory followed by a transformation or measurement from another subtheory. I then define the dimension of a sub-theory as follows.

 \begin{definition} 
 	
 	A sub-theory  $(\mathcal{P},\mathcal{M}, \mathcal{T}, p)$ of an operational theory is \textbf{d-dimensional} iff $d$ is the smallest number such that there exists a set of $d^2 - 1$  continuous parameters in $[0, 1]$ with the following properties\footnote{This is a fairly weak notion of `dimension,' in that I have not ruled out the possibility that there could be two or more disjoint subtheories classified as two-dimensional under this definition.  In other contexts it might be helpful to strengthen the definition by requiring that the subtheory of dimension two should be unique,  but this is not necessary for the derivation I give here and therefore for brevity I employ this simpler and weaker  notion of dimension.} : 
 	
 	\begin{enumerate} 
 		
 		\item Specifying the values of all $d^2 - 1$ parameters for any preparation $P \in \mathcal{P}$ fully determines the probabilities $p( M^x | P)$ for every outcome $M^x$ of every measurement $M$ in $\mathcal{M}$.

 		\item For every possible set of values of the $d^2 - 1$ parameters, there exists a preparation $P \in \mathcal{ P}$ described by those parameters. \footnote{This continuity assumption is stronger than I strictly need for the argument: it would suffice to make some kind of symmetry assumption, such as the assumption that if there exists a preparation described by a set of parameters $\vec{x}$ with $x_1 = p$, then there exists a preparation described by the same set of parameters except with $x_1 = 1 - p$. I have chosen to use the stronger assumption here as it is simpler and seems more natural; it is clear that both assumptions are satisfied by quantum mechanics.}

 	\end{enumerate} 
 	
 \end{definition} 
 
 Obviously, the choice to associate $d$ dimensions with $d^2 - 1$ parameters is inspired by quantum theory: a $d$-dimensional quantum state is described by a set of $d$ complex parameters, or $d^2$ continuous real parameters, minus $1$ to account for normalization. But this particular mapping is a naming convention only - I could equally well have chosen to say that a  subtheory described by a set of $d$ parameters is $d$-dimensional, which might have seemed the natural choice if it were not for the example of quantum theory. And therefore this definition does not take for granted any particular structure for the state space of the subtheory, other than the continuity implied by the second property. 
 
 \vspace{2mm} 
 
 The fine-grained uncertainty relation also applies only for pairs of orthogonal measurements, and therefore we require a notion of `orthogonal measurement' for the operational framework: 
 
 \begin{definition} 
 	
 	In some subtheory $(\mathcal{P},\mathcal{M}, \mathcal{T}, p)$ of an operational theory, two measurements $M_1, M_2 \in \mathcal{M}$ are \textbf{orthogonal} iff given an arbitrary unknown preparation $P$,  the set of probabilities $\{ p(M^x_1 | P) \}$ and  $\{ p(M^x_2 | P) \}$ are independent. 
 	
 \end{definition} 
 
 This definition implies that given a preparation $P$, and any set of $q$ orthogonal  measurements $\{ M_i \}$ each having two outcomes $M_i^0, M_i^1$, we need $q$ independent parameters to specify the set of $q$ probabilities $p(M_i^0| M_i, P)$. Therefore I can set $q$ out of the $d^2 - 1$ parameters needed to specify the full set of outcome probabilities to be equal to the   $q$ probabilities $p(M_i^0| M_i, P)$. In particular,  for a two-dimensional system, given any set of three orthogonal  measurements $\{ M_1, M_2, M_3 \}$ each having two outcomes, specifying the three probabilities $p(M_1^0 | M_1, P), p(M_2^0 | M_2, P), p(M_3^0 | M_3, P)$ is sufficient to fix the probabilities for any other measurement which may be performed after the preparation $P$.

 \subsection{theorem}
 
 Using these concepts, we have the following theorem (see appendix \ref{app3} for the proof). 
 
 \begin{theorem} \label{theorem3}
 	
 	If an operational theory obeys the operational Choi-Jamio\l{}kowski isomorphism and information causality, then given  any subtheory  $(\mathcal{P},\mathcal{M}, \mathcal{T}, p)$ of dimension two, for any preparation $P \in \mathcal{P}$ and any pair of orthogonal measurements $M_1, M_2 \in \mathcal{M}$, and any two outcomes $M_1^m$, $M_2^n$ of the measurements $M_1, M_2$, we must have:

 	\[  p(  M_1^m | M) + p( M_2^n | M')  \leq 1 + \frac{1}{\sqrt{2}} \]

 \end{theorem}

 \section{Discussion \label{discussion}}

 I have suggested that the study of intratheoretic causal structures offers us a  way of rethinking the structural assumptions built into our theories in a way that is independent of any particular ontology.  In particular, I have focused here on  a particularly stubborn ontological assumption which is deeply woven into  both classical and quantum physics: the idea that information about the past must be carried into the future by a mediating state.

 Although it is still very common to assume that temporal correlations must be mediated, we  do not usually suppose that the non-local multipartite correlations in quantum mechanics are mediated by an intervening process, not least because any such process would either have to travel faster than light or go backwards in time. Thus by considering an intratheoretic causal structure for quantum mechanics in which the equivalence between multipartite correlations and temporal correlations is treated as fundamental, we are moving towards a picture of quantum mechanics where temporal correlations are \emph{not} mediated. Therefore the structure set up here offers a way of thinking about quantum mechanics where neither the quantum state nor any other sort of state is a fundamental object. 
 
 In particular, I have demonstrated that several of the characteristic properties of the quantum state can in fact be derived from features of non-local correlations without appeal to any concept of state. For example, theorem \ref{theorem2} shows that using the operational Choi-Jamio\l{}kowski isomorphism, the fact that quantum states exhibit preparation contextuality can be derived from the fact that we live in a world which is non-local but non-signalling. Intuitions will vary, of course, but to me non-locality and no-signalling seem like very fundamental features of reality, and therefore I find it plausible to regard preparation contextuality as a consequence of non-locality and no-signalling. Thus preparation contextuality need not be regarded as a fundamental property of some entity known as the quantum state - it can be thought of as a behaviour that is in fact caused by very general facts about space and time together with the equivalence between multipartite correlations and temporal correlations. 
 
 The same goes for the other  derivations presented here. There is no need to postulate a quantum state to be the bearer of the properties of `no-broadcasting', 'interference,' and `obeying uncertainty relations,' since these features can be regarded as consequences of constraints on non-local correlations. Thus these derivations point towards a picture of quantum mechanics where the state is not a fundamental entity but instead emerges from deeper properties of the theory. There do exist several specific proposals for the ontology of quantum mechanics which   eliminate states from the picture - particularly those approaches where the ontology consists entirely of  events, such as the GRW flash ontology~\cite{Tumulka2006}~\cite{Bell1985} or Kent's solution to the Lorentzian quantum reality problem \cite{Kent2013} - but by  couching the discussion in terms of intratheoretic causal structure it is possible to understand some general consequences of the elimination of states without committing to specific ontological details.  
 
 I will finish by reinforcing that the possibility of a stateless ontology for quantum mechanics has significant consequences for physics outside quantum foundations, because the attitude that we take toward the quantum state has a strong influence on the way we are likely to proceed when working on extensions of the theory, such as quantum field theory and quantum gravity. If we think of the quantum state as the fundamental object of quantum theory, it is natural to deal with gravity by writing down quantum states of spacetime - that is, by 'quantizing gravity,' as is done in most of the mainstream approaches to quantum gravity, including covariant quantum gravity, canonical quantum gravity and string theory. On the other hand, if we think that the quantum state is actually just a codification of relations between events or some other ontological substratum, it seems more natural to regard spacetime itself as emerging from the underlying substratrum, as  in causal set theory. Which direction we take will depend to a large extent on the beliefs we hold, implicitly or explicitly, about the underlying intratheoretic causal structure of quantum mechanics, and so exploring new structures where states are  not fundamental may shed new light on outstanding problems in quantum field theory and quantum gravity.  

%%%%%%%%%%%%%%%%%%%%%%%%%%%%%%%%%%%%%%%%%%

 \appendix
\subsection{Proof of theorem \ref{theorem1} \label{app1}}

\begin{proof} 
	
	Consider any operational theory  $(\mathcal{P},\mathcal{M}, \mathcal{T}, p)$ which obeys the operational Choi-Jamio\l{}kowski isomorphism. Suppose  the theory   exhibits Bell-nonlocality. Then for some joint preparation  $P_{AB}$  and some pair of measurements  $M_A^0, M_A^1$ on $S_A$ and some pair of measurements $ M_B^0, M_B^1$ on $S_B$, the CHSH quantity satisfies: 
	
	\[ \mathscr{B}_{AB}(P_{AB}, M_A^0, M_A^1, M_B^0, M_B^1) > 2 \] From the operational Choi-Jamio\l{}kowski isomorphism, there exist ensemble preparations $P^0, P^1$ and  a transformation $T$ which give rise to the same probability distributions, and therefore 
	\[ \mathscr{B}_{AB}( P^0, P^1, T, M_B^0, M_B^1) > 2 \]  
	
	By stipulation (recall section \ref{ot}), I can always combine the transformation $T$ with the measurements $P^0, P^1$ to produce new measurements $N^0, N^1$, thus  there exist measurements $N^0, N^1$ such that   $\mathscr{B}_{AB}( P^0, P^1, \mathbb{I}, N^0, N^1) > 2$.

	\vspace{2mm} 
	
	Now suppose the theory $(\mathcal{P},\mathcal{M}, \mathcal{T}, p)$  has a universal broadcasting map, i.e.  a channel   which takes any given operational state  to  two perfect copies of the same operational state. 
	This channel may be represented as two copies of the identity transformation $\mathbb{I}$ acting on the input state. 
	
	Let us  apply this universal  broadcasting map to the result of an ensemble preparation $P$, producing two copies of the input state on two systems $S_B, S_C$. I may then apply measurements 
	$N_B, N_C$, giving rise to a joint distribution  $p_{P ; \mathbb{I} ; N_B}$ over the outcome of the ensemble preparation and the outcome of the measurement $N_B$, and another joint distribution   $p_{P ; \mathbb{I} ; N_C}$ over the outcome of the ensemble preparation and the outcome of the measurement $N_C$, and hence a joint distribution $p_{P ; N_B, N_C}$ over the outcome of the ensemble preparation and the two outcomes of the two measurements. 
	
	From the operational Choi-Jamio\l{}kowski isomorphism, it follows that there exists a joint preparation $P_{ABC}$ of three systems $S_A, S_B, S_C$ and a measurement $N_A$ on system $S_A$ such that if I perform $P_{ABC}$ and then the measurements $N_A, N_B, N_C$ on the three systems respectively,   the joint distribution $p_{P_{ABC} ; N_A N_B N_C}$ over the measurement outcomes is identical to the distribution  $p_{P ; N_B, N_C}$. 
	
	Thus consider applying the universal broadcasting map to either of the two ensemble preparations $P^0, P^1$ and then choosing from the set of measurements $N^0, N^1$ for the systems $S_B, S_C$. I have that: 
	
	\[ \mathscr{B}_{AB}( P^0, P^1, \mathbb{I}, N^0, N^1
	) > 2 \]    \[ \mathscr{B}_{AC}( P^0, P^1,  \mathbb{I}, N^0, N^1  ) > 2 \] 
	
	It follows that there exists a preparation $P'_{ABC}$  such that if I perform $P_{ABC}'$ and then choose from the measurements $N^0, N^1$ for the systems $S_A, S_B, S_C$, I have: 
	
	\[ \mathscr{B}_{AB}( P_{ABC}', N_A^0, N_A^1, N_B^0, N_B^1 
	) > 2 \]  \[\mathscr{B}_{AC}( P_{ABC}', \mathbb{I}, N_A^0, N_A^1, N_C^0, N_C^1   ) > 2 \]
	
	And therefore: 
	\[  \mathscr{B}_{AB}( P_{ABC}', N_A^0, N_A^1, N^0, N^1 
	)^2 + \mathscr{B}_{AC}( P_{ABC}', \mathbb{I}, N_A^0, N_A^1, N^0, N^1   ) ^2 > 8 \] 
	
	Thus the monogamy of correlations is necessarily violated in the theory $(\mathcal{P},\mathcal{M}, \mathcal{T}, p)$.

\end{proof} 

\subsection{Proof of theorem \ref{theorem2} \label{app2}}

\begin{proof}

	Consider an operational theory   $(\mathcal{P},\mathcal{M}, \mathcal{T}, p)$ which obeys the operational Choi-Jamio\l{}kowski isomorphism and the no-signalling principle. 
	
	Within such a theory, suppose we perform a joint preparation $P_{AB}$ for  two distinct systems $A, B$, then perform one measurement from a set of $n$ measurements  $\{ M_i\}$ on system $A$, and a fixed measurement $F_b$ on system $B$.

	It follows from the operational Choi-Jamio\l{}kowski isomorphism that there exists  a set of $n$ ensemble preparations $\{ P_i \}$ on a system $S$ and a channel $T$ such that if we perform the preparation $P_i$ on $S$, then input $S$  to the channel $T$, then perform measurement  $F_b$ on the product of the channel, each of the resulting distributions  $p_{ P_i ; T;  F_b}  $ is identical to the corresponding  distribution $p_{ P_{AB}; M_i F_b}$. Thus each $P_i$ is associated with a probability distribution $p^i(x)$ such that $p^i(x)  = p^A(M^x_i | M_i, P_{AB} )$.

	Since the theory is no-signalling, there is no choice of measurement $F_b$ on system $B$ which gives any information about which measurement  $M_i$ was performed on system $A$ in the former scenario, and therefore   there is no choice of measurement $F_b$ which gives any information about which ensemble preparation $P_i$ was chosen in the latter scenario. This means that any one of the preparation procedures in  $\{ P_i \}$, followed by the transformation $T_{AB}$, gives rise to the same operational state, which I denote by $\chi$.  I will henceforth denote by $P_i'$ the  composite preparation procedure obtained by performing the procedure $P_i$ followed by the transformation $T_{AB}$.

	\vspace{2mm}
	
	Let us now write down an ontological model for this operational theory.
	
	\vspace{2mm} 
	
	Each ensemble preparation $P_i$ is associated with a set of preparations $P_i^x$ each being chosen with probability $p^i(x)$. 
	Each preparation $P_i^x$ is associated with a probability distribution $\mu( \lambda |P_i^x) $ in the ontological model, and the result of performing a preparation $P^i_x$ and then applying the transformation $T$ is also associated with a probability distribution, $\mu( \lambda |T,  P^i_x) $\footnote{Note that I do not assume that there exists a unique column-stochastic matrix in the ontological model which represents the transformation $T$ - if there is more than one such transformation, I simply use any one which produces the correct statistics for this particular scenario - and therefore I need not assume that the theory is transformation contextual.}.    Thus each of the ensemble preparations $P_i \in \{ P_i \}$  gives rise to a representation of $\chi$ as a probability distribution over ontic states: 
	
	\[p( \lambda |  \chi ) =  \sum_x p^i(x)  \mu( \lambda |T, P^i_x)   \] 
	
	\[ \hspace{13mm} =   \sum_x p^A(M^x_i | M_i, P_{AB} ) \mu( \lambda |T,  P^i_x)  \]

	For any $M_i \in \{ M_i\}$,  the joint probability distribution $p_{P_{AB} ; M_i ; F_b}$ over the outcomes of the measurements on the systems $A, B$ after the preparation $P_{AB}$ can   be written as follows\footnote{Note that I do not assume that there exists a unique set of response functions $p(F^y_b | F_b, \lambda)$ for the measurement $F_b$ - if there is more than one such response function, we can simply use any one which produces the correct statistics for this particular scenario - and therefore I need not assume that the theory is measurement non-contextual.}:
	
	\begin{multline}  p(M_i^x, F_b^y)_{P_{AB}, M_i, F_b}=   p^A(M^x_i | M_i, P_{AB} ) p(F^y_b | M^x_i, M_i, F_b, P_{AB}) \\
	\hspace{33mm}  =  p^A(M^x_i | M_i, P_{AB} ) p(F^y_b | T, P^i_x) \\  
	\hspace{33mm}   = p^A(M^x_i | M_i, P_{AB} ) \sum_{ \lambda}  \xi(F^y_b | F_b, \lambda ) p( \lambda |T,  P^i_x)  \\ \end{multline}

	Now suppose the operational theory    $(\mathcal{P},\mathcal{M}, \mathcal{T}, p)$ is not preparation contextual. This means that it is possible to choose the ontological model in such a way that all representations of $\chi$ are identical, so there exists a unique probability distribution $p(\lambda)$ such that for any $i  \in \{ 1, 2, ... n \}$ I have: 
	
	\[  \sum_x p^A(M^x_i | M_i , P_{AB}) p( \lambda |T, P^i_x) =  p (\lambda ) \]  
	
	\vspace{2mm}

	Now for any $i \in \{1, 2 ... n\} $ let us define: \[ \xi(M^{x}_i | M_i, \lambda) : = \frac{p^A(M^{x}_i | M_i, P_{AB}) p( \lambda |T, P^i_x)}{  p(\lambda) } \]
	
	Note that  	$ \forall i \quad \sum_{x}\xi(M^{x}_i | M_i \lambda) = 
	\sum_x \frac{p^A(M^{x}_i | M_i, P_{AB} ) p( \lambda |T, P^i_x) }{  p(\lambda) } = \frac{p(\lambda) }{  p(\lambda) } = 1$, and since every $\xi(M^{x}_i | M_i, \lambda) $ is nonnegative, it follows that  $\forall i, x \quad \xi(M^{x}_i | M_i , \lambda)   \in [0,1] $. These are the sufficient conditions for a function to be a valid response function of an ontological model, so $\xi(M^{x}_c | M_a, \lambda) $ is a valid response function. 
	
	\vspace{2mm}

	The joint probability distribution over the outcomes of the measurements on the systems $A, B$ after the preparation $P_{AB}$ can then be written as follows: 
	
	\begin{multline}  p(M_i^x, F_b^y)_{P_{AB}, M_i, F_b}   =  \sum_ { \lambda}  \xi(F^y_b | F_b, \lambda ) \xi(M^{x}_i | M_i, \lambda) p( \lambda) \\ \end{multline}

	Thus any joint probability distributions within the theory $(\mathcal{P},\mathcal{M}, \mathcal{T}, p)$ are factorizable,  and it is a well-known result\cite{Bell} that measurements described by factorizable probability distributions cannot satisfy $\mathscr{B}_{AB} > 2$, so the theory $(\mathcal{P},\mathcal{M}, \mathcal{T}, p)$  cannot exhibit Bell nonlocality.

\end{proof}

\subsection{Proof of theorem \ref{theorem3} \label{app3}}

\begin{proof} 
	
	Let   $M_1, M_2, M_3 \in \mathcal{P}$ be a set of three orthogonal measurements for the two-dimensional subtheory $(\mathcal{P},\mathcal{M}, \mathcal{T}, p)$, each having two outcomes which I label by $0, 1$. 
	
	\vspace{2mm} 
	
	Let $P_a \in \mathcal{P}$ be a preparation which maximizes $ p(0 | M_1, P_a )   + p(0 | M_2, P_a)$, with the maximization performed over all preparations in  $\mathcal{P}$. Let us say that $ p(0 | M_1, P_a )  = i$, $p(0 | M_2, P_a)= j$ and $p(0 | M_3, P_a) = k$  Since I can choose $p(0 | M_1 ,P ), p(0 | M_2, P),  p(0 | M_2, P)$ to be the set of three parameters which suffice to fix the measurement outcomes for any other  measurement performed after $P$, and by definition, for any valid  set of values of these  parameters, there exists a preparation $P \in \mathcal{ P}$ described by those parameters, it follows that there exist preparations $P_b , P_c, P_d$ with outcome probabilities as follows: 
	
	\vspace{2mm} 
	\begin{tabular} { c|  c | c |  c | c}
		& $ P_a$& $ P_b$& $ P_c$& $ P_d$	\\ 
		$	p( 0 | M_1 )$ & $ i $ & $ 1 - i$ & $i$ & $ 1 - i$ \\
		$	p( 0 | M_2 )$& $ j $ & $ 1 - j$ & $ 1 - j$ & $ j$ \\
		$ p( 0 | M_3 ) $& $ k$ & $  k $ & $  k$ & $ k$ \\
	\end{tabular} 
	\vspace{2mm}

	Suppose Alice performs either preparation $P_a$ or $P_b$, chosen uniformly at random, and then sends the resulting system to Bob. If Bob then performs the measurement $M_1$, he will obtain the outcome $0$ with probability $\frac{1}{2} $ and if Bob then performs the measurement  $M_2$, he will obtain the outcome $0$ with probability $\frac{1}{2} $.  	Likewise, if Alice performs $P_c$ or $P_d$ chosen uniformly at random and then sends the   system to Bob, if Bob performs the measurement $M_1$ Bob will   obtain     the outcome $0$ with probability $\frac{1}{2} $ and if Bob then performs the measurement  $M_2$, he will obtain the outcome $0$ with probability $\frac{1}{2} $. 	Clearly the probabilities for the outcomes of measurement $M_3$ are also the same for both decompositions. 	Since the probabilities $p(0 | M_1 , P), p(0 | M_2, P),  p(0 | M_2, P)$  suffice to fix the measurement outcomes for any other  measurement performed after $P$, this implies that these two probabilistic preparations give rise to the same measurement statistics for any subsequent measurements, and hence  both decompositions produce the same operational state, which I denote by $\chi$. 
	\vspace{2mm}
	
	It follows from the operational Choi-Jamio\l{}kowski isomorphism that there exists a preparation $P_{AB}$ and a measurement $S_0$ such that when $S_0$ is performed on one subsystem $A$ of a bipartite system $AB$ jointly prepared using preparation $P_{AB}$ with probability $\frac{1}{2}$ I obtain outcome $0$ and the resulting operational state of subsystem $B$ is the same as the operational state produced by preparation $P_a$, and with probability $\frac{1}{2}$ I obtain outcome $1$ and the resulting operational state of subsystem $B$ is the same as the operational state produced by preparation $P_b$. Likewise, there exists a measurement $S_1$ defined similarly for preparations $P_c, P_d$.

	\vspace{2mm}

	Now consider  a CHSH game in which Alice and Bob are each given input bits $c, d $ and are required to produce outcomes $a, b$ such that $a \oplus b =  c d $. Alice and Bob may employ the following strategy: 
	
	\begin{enumerate} 
		
		\item Before the start of the game, Alice and Bob perform the preparation $P_{AB}$, and then Alice takes subsystem $A$ and Bob takes subsystem $B$.

		\item When Alice is given input $c \in \{ 0, 1\}$, she performs measurement $S_c$ and then returns her measurement outcome. 
		
		\item When Bob is given input $d \in \{ 0, 1\}$, he performs measurement $M_{d + 1}$ and then returns his measurement outcome. 
		
	\end{enumerate} 
	
	If the inputs are chosen uniformly at random, the probability that Alice and Bob win the game using this strategy is: 
	
	\[\frac{1}{4} \sum_{c, d, a} p(a|c) p(x(a c d) |a c d)\]
	
	where  $p(a | s)$  the probability that Alice obtains outcome $a$ when she performs measurement $s$, 	$x(a c d)$ is defined such that  for any $a, c, d$, $a \oplus x(a, c, d)= c d $,  and  $p( q |a c d )$ is the probability that Bob obtains outcome $q$ when Alice performs measurement $S_c$ and obtains outcome $a$ and Bob performs measurement $M_{d + 1}$.

	\vspace{2mm} 
	
	$S_0$ and $S_1$ are defined such that   $\forall a \ \forall s \;  p(a|s)  = \frac{1}{2}$.  	Moreover, by the definition of $S_0$, if Alice performs measurement $S_0$ and obtains outcome $0$, the resulting probabilities $p(q | 0, 0, d)$  for the outcomes of Bob's measurement $M_{d + 1}$ are the same as if Alice were to simply perform preparation $P_a$ and hand the prepared system over to Bob. That is, $p(0 | 0,0,0) = i$ and $p(0 | 0,0,1) = j$. Similar relationships hold for other possible values of $a$ and $c$, and using the table above, it is straightforward to obtain: 
	
	\[  \forall a, c \quad \sum_{d = 1}^2 p(x(a, c, d) | a, c, d) = i + j \]

	Hence the probability that Alice and Bob win the game using this strategy is $\frac{1}{2} (i + j)$. 
	
	Ex hypothesi the operational theory satisfies information causality. It is known that information causality implies the Tsirelson bound\cite{Pawlowski}, and the Tsirelson bound implies that the maximum probability of winning this CHSH game is $1 + \frac{1}{\sqrt{2}}$; hence  $i + j \leq 1 + \frac{1}{\sqrt{2}}$.
	
	Since we chose $P_a$ to be the preparation which maximizes $p(0 | M_1, P_a )   + p(0 | M_2, P_a)$, it follows that for any preparation $P$, $p(0 | M_1, P )  + p(0 | M_2, P)  \leq 1 + \frac{1}{\sqrt{2}}$. 
	
	\vspace{2mm} 
	
	Then note that for any  $m \in \{0, 1\}, n \in \{ 0, 1\}$ I can make a similar argument by starting from the requirement that  $P_a$ is the state which maximizes  $p(m | 0 ) + p(n | 1)$, and hence:

\end{proof}

\bibliography{newlibrary11}{}
\bibliographystyle{unsrt}

\end{document}